\documentclass[sigconf]{acmart}

\usepackage{booktabs}
\usepackage{multirow}
\usepackage{multicol}
\usepackage{bbding}
\usepackage{threeparttable}

\usepackage{algorithm}
\usepackage{algorithmic}

\AtBeginDocument{%
  }

\copyrightyear{2022}
\acmYear{2022}
\setcopyright{acmcopyright}
\acmConference[DDAM '22] {Proceedings of the 1st International Workshop on Deepfake Detection for Audio Multimedia}{October 14, 2022}{Lisboa, Portugal.}
\acmBooktitle{Proceedings of the 1st International Workshop on Deepfake Detection for Audio Multimedia (DDAM '22), October 14, 2022, Lisboa, Portugal}
\acmPrice{15.00}
\acmISBN{978-1-4503-9496-3/22/10}
\acmDOI{10.1145/3552466.3556527}

\begin{document}

\title{Deep Spectro-temporal Artifacts for Detecting Synthesized Speech}

\author{Xiaohui Liu}
\email{liuxiaohui2021@tju.edu.cn}
\affiliation{%
  \institution{Tianjin University}
  \city{Tianjin}
  \country{China}
}

\author{Meng Liu}
\affiliation{%
  \institution{Tianjin University}
  \city{Tianjin}
  \country{China}
}

\author{Lin Zhang}
\affiliation{%
  \institution{National Institute of Informatics}
  \city{Tokyo}
  \country{Japan}
}

\author{Linjuan Zhang}
\affiliation{%
  \institution{Taiyuan University of Technology}
  \city{Taiyuan}
  \country{China}
}

\author{Chang Zeng}
\affiliation{%
  \institution{National Institute of Informatics}
  \city{Tokyo}
  \country{Japan}
}

\author{Kai Li}
\affiliation{%
  \institution{Japan Advanced Institute of Science and Technology}
  \city{Nomi, Ishikawa}
  \country{Japan}
}

\author{Nan Li}
\affiliation{%
  \institution{Tianjin University}
  \city{Tianjin}
  \country{China}
}

\author{Kong Aik Lee$^{*}$}
\affiliation{%
  \institution{Institute for Infocomm Research, A$^{\star}$STAR}
  \country{Singapore}
}

\author{Longbiao Wang$^{*}$}
\affiliation{%
  \institution{Tianjin University}
  \city{Tianjin}
  \country{China}
}

\author{Jianwu Dang}
\affiliation{%
  \institution{Japan Advanced Institute of Science and Technology}
  \city{Nomi, Ishikawa}
  \country{Japan}
}


\renewcommand{\shortauthors}{Xiaohui Liu et al.}

\begin{abstract}
The Audio Deep Synthesis Detection (ADD) Challenge has been held to detect generated human-like speech. With our submitted system, this paper provides an overall assessment of track 1 (Low-quality Fake Audio Detection) and track 2 (Partially Fake Audio Detection). In this paper, spectro-temporal artifacts were detected using raw temporal signals, spectral features, as well as deep embedding features. To address track 1, low-quality data augmentation, domain adaptation via finetuning, and various complementary feature information fusion were aggregated in our system. Furthermore, we analyzed the clustering characteristics of subsystems with different features by visualization method and explained the effectiveness of our proposed greedy fusion strategy. As for track 2, frame transition and smoothing were detected using self-supervised learning structure to capture the manipulation of PF attacks in the time domain. We ranked 4th and 5th in track 1 and track 2, respectively. 
\end{abstract}



\maketitle

\section{Introduction}
\label{sec:intro}

Automatic Speaker Verification (ASV) has been widely used in security, human-computer interaction, and other fields. However, similar to other biometric authentication methods, ASV systems are vulnerable to attacks from speech generation technology \cite{wu2015spoofing}. Synthesized speech generated by deep learning algorithms is even difficult to be distinguished by human ears, which poses a great threat to ASV system. Thus, it is of great significance to build a countermeasure (CM) system that can effectively distinguish spoof speech from genuine speech.

\begin{table*}[htbp]
\centering
\caption{Database details of the LF and PF track. Duration, Mean, Max, and Min are all measured in second.}
\label{tab:data}
\begin{tabular}{@{}ccccccccc@{}}
\toprule
\textbf{Track}                   & \textbf{Subset}      & \textbf{Total} & \textbf{Genuine} & \textbf{Fake} & \textbf{Duration (s)} & \textbf{Mean (s)} & \textbf{Max (s)} & \textbf{Min (s)} \\ \midrule
\multirow{2}{*}{\textbf{}} & Training    & 27,084          & 3,012               & 24,072          & 85,355.51           & 3.15             & 60.01         & 0.86         \\
                                 & Development & 28,324          & 2,307               & 26,017          & 89,567.78           & 3.16             & 60.01         & 0.86         \\ \midrule
\multirow{2}{*}{\textbf{LF}}     & Adaptation1  & 1,000           & 300                & 700            & 3,627.24            & 3.63             & 60.01         & 1.13        \\
                                 & Test1        & 109,199         & -                  & -              & 383,687.88          & 3.51             & 158.14        & 0.26        \\ \midrule
\multirow{3}{*}{\textbf{PF}}     & Adaptation2  & 1,052               & 0                   &    1,052            & 4,568.56                  & 4.34                 &  13.42            & 1.30             \\
                                 & Adaptation2$^\star$  & 2,104               & 1,052                   &    1,052            &      -             &         -         &           -   &       -       \\
                                 & Test2        & 100,625               & -                   & -               & 77,625.32                  & 5.94                 & 68.46             & 1.53              \\ 
                                 \bottomrule
\end{tabular}
\vspace*{-2mm}
\end{table*}

Previous ASVspoof challenges \cite{Wu2014, Kinnunen2017, Nautsch2021, asv2021_2} have played a key role in fostering spoofed speech detection research but ignored three practical scenarios: 1) diverse background noises and disturbances in the fake audios, 2) partial spoofed segments in a real speech audio, and 3) new algorithms of speech synthesis and voice conversion are proposed rapidly. Besides, the existing anti-spoofing databases are mainly in English.

To address these problems, the first Audio Deep Synthesis Detection Challenge (ADD 2022) \cite{add2022} focuses on three different tracks using a Mandarin corpus: Low-quality Fake Audio Detection (LF), Partially Fake Audio Detection (PF), and Audio Fake Game (FG). LF task consists of genuine and entirely fake utterances generated using the text-to-speech (TTS) and voice conversion (VC) algorithms but contains various background noises and disturbances. PF task comprises genuine and partially fake utterances generated by combining the original genuine utterances with real or synthesized utterances. According to the challenge plan \cite{add2022}, all three tasks share the same training and development datasets but different adaptation and test datasets. Therefore, domain mismatch between training and test is a severe challenge for all three tracks.  

This paper describes our systems submitted to the ADD 2022. We focused on the LF and PF track. For the LF track, the proposed low-quality data augmentation, domain adaptation via finetuning, and greedy fusion of various complementary features were respectively aggregated in our system. Then we submitted a primary system for track 1. For the PF track, we conducted several popular Self-supervised learning (SSL) models with two Bidirectional Long Short-Term Memory (Bi-LSTM) layers \cite{2005Framewise} and one fully-connected layer as our basic system. We fine-tuned all models with adaptation set to reduce the mismatch between train set and test set. We submitted a single system for track 2.

The rest of the paper is organized as follows: Section 2 presents the details about ADD 2022 challenge and data. Section 3 describes our single systems and their setups. Section 4 presents and analyzes experimental results. In Section 5, we summarize our conclusions.

\section{Task Description and Data}
\label{sec:data}

The data for the challenge consists of training, development, adaptation, and test sets. Utterances in both training and development sets are noiseless, where genuine utterances are selected from clean AISHELL-3 \cite{2020AISHELL} corpus, and fake utterances are generated by mainstream speech synthesis and voice conversion systems based on AISHELL-3. Data for the LF track and the PF track is distributed as 16 kHz, 16 bits per sample WAV files. The information of datasets is summarized in Table \ref{tab:data}. Since the organizers only provide fake audio for the adaptation set of track 2, we randomly select 1052 (same number as fake) genuine utterances from the training and development set to make the balance between fake and genuine, and we call it as Adp2$^\star$, which will be used to compare SSL models. 

For the LF track, the adaptation and test sets are composed of genuine and fully fake utterances, but the test set is more complicated. Since provided utterances in training and development sets are noiseless, this track requires developing systems robust to the noisy environments under the scenario of mismatch between training and testing data. 

PF track is a new topic proposed in 2021 \cite{Zhang2021PartialSpoof, Yi2021halftruth}. As the name suggests, the spoofed utterances in this scenario are partially faked. Thus genuine parts that exist in the partially audio utterances can degrade the performance of CM systems, making this track more challenging. Besides, since fake utterances contain genuine or spoofed clippings, the key point to detect partially faked audio is to detect the frame transition or change in the temporal domain.

Metric for both tracks is Equal Error Rate (EER) \cite{add2022}.

\section{System Description}
\label{sec:system}

\subsection{Features}
\label{sec:features}
As shown in Table \ref{tab:feature}, we explore various features in three categories: temporal raw waveform, hand-crafted features, and deep embedding feature. The combination of these features is expected to fully capture the spectro-temporal divergences between spoof and genuine utterances.

Besides, we apply different trimming strategies of input for two different tracks. For the LF track, an unified duration of 3s has been applied for feature extraction. For the PF track, as entire utterance information is of significant importance, the entire utterance is regarded as input \footnote{Actually, upper limitation duration is 15s because of computation limitation}. Average Pooling layer is employed to handle inputs with varied duration. 

\begin{table*}[htbp]
\centering
\caption{Details of features.}
\label{tab:feature}
\begin{tabular}{@{}ccccc@{}}
\toprule
\multicolumn{1}{c}{Feature Categories}     & Feature Name           & Dimension & Resolution & Domain                                                                                   \\ \midrule
\multicolumn{1}{c}{Raw Feature}            & Raw wave               & 64,000       & -                   & Temporal                                                                                 \\ \midrule
\multirow{6}{*}{Hand-crafted Feature}   & Magnitude spectrogram  & 513       & Linear-scale        & \multirow{4}{*}{\begin{tabular}[c]{@{}c@{}}Spectro-temporal \\ (Magnitude)\end{tabular}} \\ 
                                             & Fbank                  & 80        & Mel-scale           &                                                                                          \\ 
                                             & MFCC                   & 23        & Mel-scale           &                                                                                          \\ 
                                             & CQT gram               & 96        & Variable-scale      &                                                                                          \\ \cline{2-5} 
                                             & Phase spectrogram      & 256       & Linear-scale        & \multirow{2}{*}{\begin{tabular}[c]{@{}c@{}}Spectro-temporal\\ (Phase)\end{tabular}}      \\ 
                                             & MRP                    & 120       & Mel-scale           &                                                                                          \\ \midrule
\multicolumn{1}{l}{Deep Embedding Feature} & SSL  &  1,024         & -                   & Channel                                                                                 \\ \midrule
\end{tabular}
\vspace*{-2mm}
\end{table*}

\subsubsection{Temporal raw feature}
\label{sssec:trf}

\ 
\newline
\noindent
Considerable evidence \cite{tak2021end} shows that avoiding the use of hand-crafted features with end-to-end architecture may improve the performance of anti-spoofing systems. Therefore, we utilize raw waveform as input following Sinc filters \cite{2019SincNet}. 

\subsubsection{Hand-crafted features}
\label{sssec:spf}
\ 
\newline
\noindent 
Hand-crafted features are most commonly used in anti-spoofing, since they contain specific knowledge. In this subsection, the following applied or designed features are all online GPU features, which accelerate the training speed.

\textbf{Spectrogram}. The log-Spectrogram \cite{li2021replay} is extracted with 50 ms frame length, 25 ms frame shift, 1024 FFT point and hamming window. Both magnitude and phase spectrogram are extracted, relatively.

\textbf{CQT} (Constant Q Transform) gram. The length of the filter window is changeable according to different frequencies. The log-CQT gram is extracted with fmin of 32.7 Hz, Hanning window, 84 bins with 12 bins per octave \footnote{https://kinwaicheuk.github.io/nnAudio/index.html}.

\textbf{Fbank} (Mel Filterbanks) and \textbf{MFCC} (Mel Frequency Cepstral Coefficient) \cite{murty2005combining}. Log-Fbank and MFCC we used are extracted with the same configuration as the spectrogram.

{\bf MRP} (Mel Relative Phase). Relative phase, Mel filterbanks, and normalization are calculated \footnote{https://github.com/DanielMengLiu/Mel-Relative-Phase} according to \cite{liu2021replay}.

\subsubsection{Deep embedding feature}
\label{sssec:dmf}

\ 
\newline
\noindent 
Recently, Self-Supervised Learning (SSL) shows excellent performance in the common fully spoofed scenario in ASVspoof challenge \cite{wang2021ssl, chen2021wavlm, 2022Automatic, 9747605}. Thus, in this study, we apply several SSL models to extract deep embedding feature for PF track: Wav2Vec 2.0 Large \cite{BaevskiZMA20_w2v2}, XLSR-53 \cite{conneau2020xlsr} and WavLM Large \cite{chen2021wavlm}. Wav2Vec 2.0 Large (denotes as W2V2-Large) and XLSR-53 (denotes as XLSR) are the best two SSL models for ASVspoof in a recent study \cite{wang2021ssl}, and WavLM Large (denotes as WavLM-Large) shows superior performance in the leaderboard of SUPERB \cite{yang21c_interspeech_superb}.
SSL model will be finetuned with ADD data. And the final representation is weighted sum through all hidden features from the transformer layers in SSL with trainable weights following SUPERB \cite{yang21c_interspeech_superb}.

\subsection{Data augmentation}
\label{sec:da}

As test sets of ADD 2022 include unseen genuine and fake utterances which are not present in train and development data, it is essential to develop CMs that are robust to out-of-domain data. Data augmentation strategy is efficient to improve the performance of anti-spoofing systems on cross-dataset in previous works \cite{empirical,ur,stc,biometric, chen21b_asvspoof, das21_asvspoof, kang21b_asvspoof, fu2022fastaudio, zhang2020adversarial, yang2019sjtu, cai2019dku}. Thus, we design low-quality data augmentation strategy to address the unseen genuine and fake utterances. 

Since the LF track focuses on robustness against diverse background noises and disturbances effects, we apply background noise and reverberation to the training set. Background noise including music, noise, gaussian white noise derived from MUSAN\footnote{http://www.openslr.org/17/}, and reverberation is based on random simulated reverberation derived from RIR datasets\footnote{https://www.openslr.org/28/}. Besides, 1/3 fade in and 1/3 fade out of each audio have also been added to the data augmentation strategy, which simulate the real-world volume disturbance. 


\subsection{Deep detection network}
\label{ssec:sys_LF}



Four types of deep neural network are explored in this paper: SE-Res2Net50 \cite{li2021replay} based on Conv2d, RawNet2 \cite{tak2021end} based on Conv1d, ResNet-TCN \cite{abedi2021improving} based on the combination of Conv2d and Conv1d and BLSTM. Conv1d and Bi-LSTM are expected to capture long-term temporal artifacts, while Conv2d is expected to capture abundant spectral artifacts.

\subsubsection{SE-Res2Net50}
\label{sssec:se-resnet50}
\ 
\newline
\noindent 
SE-Res2Net50 has shown great performance in the field of anti-spoofing \cite{li2021replay} which integrates the squeeze-and-excitation (SE) block \cite{hu2018squeeze} onto the Res2Block \cite{gao2019res2net} backbone. During the training stage, we use binary cross-entropy as loss function, and Adam \cite{2014Adam} as the optimizer to train a model. The parameters of optimizer are set with $\beta_1 = 0.9$, $\beta_1 = 0.98$ and weight decay $10^{-9}$. During training, we change the learning rate according to the formula 3 in \cite{vaswani2017attention} that increases linearly for the first 1000 warmup-steps and initialized as $3 \times 10^{-4}$. We train models with 20 epochs and choose the lowest EER on the adaptation set to evaluate on the test set. 

\subsubsection{RawNet2}
\label{sssec:RawNet2}
\ 
\newline
\noindent 
RawNet2 operates upon raw audio shows generalization ability for anti-spoofing tasks \cite{tak2021end}. The first layer of RawNet2 is sinc layer with filter lengths as 129 samples. Then, six residual blocks are used to extract frame-level representations. In an effort to obtain more discriminative representations, filter-wise feature map scaling and attention mechanism are applied to each residual block output. Next, we apply Gate Recurrent Units (GRU) \cite{2014Learning} with 1024 hidden nodes to aggregate utterance-level representation. Finally, a fully connected layer is applied after the GRU layer. A softmax function is used in the output layer to produce two-class predictions: fake or genuine. The network is trained with a batch size of 32, a learning rate of 0.0001, and 100 epochs.

\subsubsection{ResNet-TCN}
\label{sssec:ResNet-TCN}
\ 
\newline
\noindent 
Temporal convolutional network (TCN) is be proven can capture long-term temporal information \cite{bai2018empirical}. Temporal convolutional blocks followed by an 18-layer residual network are stacked sequentially to act as a deep feature sequence extractor. The temporal receptive field of a standard TCN is kept fixed for all activations at a specific layer;
four branches and two layers have been designed to provide variable receptive fields to fuse short-term and long-term temporal information during feature extraction.



\subsubsection{Downstream model of SSL}
\label{sec:sys_PF}
\ 
\newline
\noindent 
We apply the powerful SSL model to extract the temporal representation and feed it into a simple downstream model to do detection. During training, the SSL model and downstream layers are jointly updated. 

The main structure of downstream following relatively better configuration on \cite{wang2021ssl} as \texttt{LGF}: SSL model + 2 Bi-LSTM (Bi-directional Long Short-Term Memory) + 1 FC (Fully-Connected layer). Moreover, to utilize the information from different transformer layers of SSL, we apply summation weight following SUPERB \cite{yang21c_interspeech_superb}. To adapt pre-trained SSL model to PF scenario in mandarin, we fine-tune this \texttt{LGF} model on (1) train + development set and (2) adp2$^\star$ set sequentially. Both entire \texttt{LGF} model and weights for different transformer layers are updated simultaneously.  

\subsection{Greedy fusion}
\label{sssec:subsubhead}

To fuse prediction from above-mentioned different subsystems, we design a greedy fusion strategy in this section as described in algorithm \ref{alg:Framwork}. Given a set $S$ involves scores and EERs produced by $N$ subsystems, we select each subsystem strategically. Firstly, we initialize the best subsystem as $S_{primary}$ that achieves the lowest EER of all subsystems. Then, we update $S_{primary}$ until the subsystem set $S$ is empty.

\begin{algorithm}[htb]
\caption{ Greedy fusion.}
\label{alg:Framwork}
\begin{algorithmic}[1] 
\REQUIRE ~~\\ 
    Set of subsystems, $S= {\textstyle \bigcup_{i=1}^{N}}S_{i}$, where $S_{i} = \left \{ score_{i}  ,EER_{i}\right  \} $;\\
    Importance factor, $\mu$;\\
\ENSURE ~~\\ 
    Primary system, $S_{primary}$;\\
    Selected subsystems, $T$;
    \STATE Get best system $S_{best}$ in $S$;
    \STATE $S_{primary}\gets S_{best}$;
    \STATE $T=T \bigcup S_{best}$;
    \STATE $S=S \backslash S_{best}$;
    \WHILE {$( S \neq \phi)$}
        \STATE $S_{before}\gets S_{primary}$;
        \STATE Get best system $S_{best}$ in $S$;
        \STATE Update $S_{primary}$ using Eq.(1);
        \STATE $S=S\backslash S_{best}$;
        \IF{$EER_{primary}\le EER_{before}$}
            \STATE $T=T \bigcup S_{best}$;        
        \ELSE
            \STATE $S_{primary}\gets S_{before}$;
        \ENDIF
    \ENDWHILE
\RETURN $S_{primary}$, $T$; 
\end{algorithmic}
\end{algorithm}

For the $i$-th iteration fusion, current primary system $S_{primary}^{(i)}$ is fused with the best unselected subsystem $S_{best}$ using an importance factor of $\mu$. $\mu$ is set to 0.9 in this work. Details are shown as the following equation:

\begin{equation}
{S_{primary}^{(i+1)}} = \mu  \cdot {S_{primary}^{(i)}} + (1-\mu)  \cdot {S_{best}}
\end{equation}

\section{Results and Discussion}

\subsection{Low-quality fake audio detection}
\label{ssec:da_LF}

\subsubsection{Results on data augmentation}
\label{sec:result_da}
\ 
\newline
\noindent 
Data augmentation strategy is adopted in the LF track for domain adaptation. To verify the effectiveness of this strategy, we firstly conduct experiments without fine-tuning on the Adp1. Experiment in each group adopts the same setup and training strategy. 

According to Table \ref{res_da}, results show that data augmentation can improve subsystems in different degrees. Subsystem based on log-Spectrogram and SE-Res2Net50 with data augmentation shows relatively 55\% improvement in adaptation set. However, it improves less on RawNet2-based systems on the adaptation set of 9\%. Since data augmentation strategy is contribute to improve the generalization ability of subsystems, we mostly adopt this strategy in the following experiments.

\begin{table}[!htp]
\centering
\caption{Comparison of data augmentation on different feature in Adp1.}
\label{res_da}
\begin{tabular}{ccccc}
\toprule
\textbf{Feature} & \textbf{Model} &  \textbf{Data Aug.} &\textbf{EER} (\%) \\
\midrule
\multirow{2}{*}{log-Spectrogram}  & \multirow{2}{*}{SE-Res2Net50} & -          & 18.69  \\
                 &              & \Checkmark          & \textbf{8.38}       \\
\midrule
\multirow{2}{*}{Raw wave} & \multirow{2}{*}{RawNet2} & -          & 13.62        \\ 
         &         & \Checkmark          & \textbf{12.38} \\
\bottomrule
\end{tabular}
\vspace*{-4mm}
\end{table}

\subsubsection{Results on domain adaptation via fine-tuning}
\label{sec:result_finetune}
\ 
\newline
\noindent 
Although data augmentation strategy is applied to training data, it couldn't remove mismatch between training data and all complex scenes ( noisy environments and unseen attacks). Thus, we further apply fine-tuning to improve performance by using the Adp1.


\begin{table}[!htp]
\centering
\caption{Results on domain adaptation via fine-tuning on Test1.}
\label{res_ft}
\begin{tabular}{ccccc}
\toprule
\textbf{Feature} & \textbf{Model} &  \textbf{Fine-tune} &\textbf{EER} (\%) \\
\midrule
\multirow{2}{*}{log-Spectrogram} & \multirow{2}{*}{SE-Res2Net50} &- &30.53  \\
                  &                   &\Checkmark  &\textbf{28.31}  \\ \hline
\multirow{2}{*}{log-CQTgram} & \multirow{2}{*}{SE-Res2Net50}                   &-  &33.96  \\
                  &                   &\Checkmark  &\textbf{30.47}  \\ \hline
\multirow{2}{*}{log-Fbank} & \multirow{2}{*}{SE-Res2Net50}                  &-  &29.70  \\
                  &                   &\Checkmark  &\textbf{28.17}  \\ \hline
\multirow{2}{*}{MRP} & \multirow{2}{*}{SE-Res2Net50}                  &-  &33.86  \\
                  &                   &\Checkmark  &\textbf{33.30}  \\ \hline
\multirow{2}{*}{Raw wave} & \multirow{2}{*}{RawNet2} &-  &43.17  \\
                  &                   &\Checkmark  &\textbf{26.86} \\
\bottomrule
\end{tabular}
\vspace*{-4mm}
\end{table}

\label{sec:results}
\begin{table*}[!ht]
\centering
\caption{Results on the LF track.}
\label{result}
\begin{threeparttable}
\begin{tabular}{@{}cccccccc@{}}
\toprule
\multirow{2}{*}{\textbf{ID}} & \multirow{2}{*}{\textbf{Feature}} & \multirow{2}{*}{\textbf{Model}} & \multirow{2}{*}{\textbf{Train Data}} & \multirow{2}{*}{\textbf{Data Aug.}} & \multirow{2}{*}{\textbf{Fine-tune}} & \multicolumn{2}{c}{\textbf{EER (\%)}} \\ \cmidrule(l){7-8} 
                             &                                   &                                 &                                      &                                    &                                    & \textbf{Adp1}     & \textbf{Test1}     \\ \midrule
A                            & log-Spectrogram                   & SE-Res2Net50                    & Train+Dev                            & Noise/Reverb/Music                 & \Checkmark                               & -                & 28.31             \\
B                           & log-CQTgram                       & SE-Res2Net50                    & Train                                & None                               & \Checkmark                                & -                & 30.47             \\
C                            & log-Fbank                         & SE-Res2Net50                    & Train                                & Noise/Reverb/Music                 & \Checkmark                                & 0.69               & 28.17            
 \\
D                            & Phase Spectrogram                               & SE-Res2Net50                    & Train                                & None                               & -                                  & -                & 39.69     
\\
E                            & MRP                               & SE-Res2Net50                    & Train                                & None                               & \Checkmark                                & -                & 33.30             \\
F                            & Raw wave                          & RawNet2                         & Train+Dev                            & Noise/Reverb/Music                 & \Checkmark                                & -                & 26.86    \\
G     &  log-Fbank        & ResNet-TCN   &   Train   &     Noise/Reverb/Music/Fade        &       \Checkmark               &      -            &        26.42          \\

\midrule
Primary                       &  G+F+C+E+B+A+D  &     [greedy fusion]                             &                                      &                                &                                    &                  &    \textbf{25.91}               \\ \bottomrule
\end{tabular}
\end{threeparttable}
\vspace*{-2mm}
\end{table*}

The shallow convolution layers of convolutional neural network are used to extract basic embeddings and the deep convolution layers are used to extract abstract embeddings. Thus, the first few layers of the pre-trained subsystem have the ability to extract the common shallow embeddings of known and unknown utterances. During fine-tuning, we freeze the initial 4 convolution layers and only update the deeper convolution layers of the pre-trained subsystem to extract abstract embeddings for unknown utterances in Test1.

Table \ref{res_ft} investigates the influence of fine-tuning strategies on different subsystems on Test1. The experiment in each group adopts the same setup and training strategy, including training data and data augmentation. The experimental results show that models trained with fine-tuning strategies can improve the detection performance for unseen noisy utterances. All SE-Res2Net50 models after fine-tuning shows slight improvement on the test set. And RawNet2 model get a significant decrease of about 37.8\%in EER.


\subsubsection{Primary system result}
\ 
\newline
\noindent 
Results for all systems on track 1 are summarized in Table \ref{result}. 

\begin{figure*}[htbp]
\centering
\includegraphics[width = 5in,height=5in ]{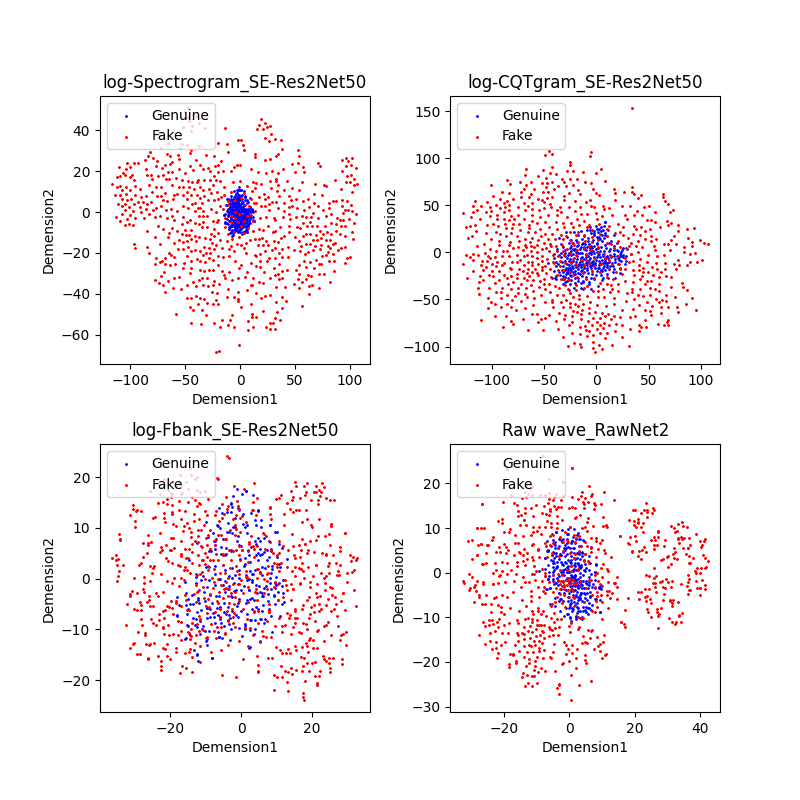}
\caption{The t-SNE visualization of the embeddings of different subsystems on the Adp1.}
\label{fig1}
\end{figure*}

Firstly, to explore the complementary information provided by different subsystems, t-SNE \cite{2008Visualizing} is applied to visualize the high-dimensional embeddings as Figure \ref{fig1} shows. We select subsystems A, B, C, F based on Table \ref{result}, including SE-Res2Net50 with log-Spectrogram, log-CQT, log-Fbank and Rawnet2 with Raw wave. We can see that all the subsystems have the ability to make genuine utterances have a more compact distribution, among which the clustering effect of subsystem A is more obvious. Subsystem F also achieves a good clustering effect of genuine utterances by adopting Rawnet2 model with Raw wave. Thus, by adopting the greedy fusion strategy, we can effectively fuse the complementary information of different subsystems and get a primary system with better performance.


Finally, we fuse seven remarkable and complementary single systems with the proposed greedy fusion strategy as our primary system. This primary system integrates spectro-temporal features of different levels and makes full use of complementary information between features, thus improving the robustness of the system against noise environments. In this way, we get EER performance of 25.91\% on the test1 set, which ranked 4th on the leader board.

\subsection{Partially fake audio detection}
\label{sec:da_PF}
This section describes the results of our submitted system for the PF track. As discussed in section \ref{sec:data}, the manipulations of PF happen in the time domain, and the artifacts mainly exist in the time domain during the processing of concatenation. Thus, temporal information is a crucial cue for CM to detect partially fake audio. Therefore, we mainly utilize SSL for the PF track. And no data augmentation is applied in this track.

\subsubsection{Results on different SSL models}
\ 
\newline
\noindent 

To compare different SSL models in the Partially Fake track, we set the learning rate as $1e-6$ following the best configuration of \cite{wang2021ssl}. Results are shown in Table \ref{tab:2_ssl}. We can see that three ssl models have similar performance on the Adp2$^{\star}$ set, while XLSR and WavLM-Large outperform W2V2-Large on the Test set. Thus, we utilize XLSR and WavLM-Large to do further exploration to select the final system. 

\begin{table}[!htp]
\centering
\caption{Comparison of different SSL model. (Adp2$^{\star}$ for Adaptation set after balance)}
\label{tab:2_ssl}
\begin{tabular}{ccccc}
\toprule
 \textbf{SSL}    & \textbf{Learning}        &  \multicolumn{2}{c}{\textbf{EER} (\%)} \\
\cmidrule(r){3-4}
\textbf{Model} & \textbf{Rate} & \textbf{Adp2}$^{\star}$ & \textbf{Test} \\
\midrule
W2V2-Large   & 1e-6                   &  33.08   & 51.76         \\
WavLM-Large & 1e-6                   &  33.56   & 45.46         \\
XLSR         & 1e-6                   &  34.32   & 44.32    \\
\bottomrule
\end{tabular}
\vspace*{-4mm}
\end{table}

\subsubsection{Results on different learning rate and fine-tuning}
\ 
\newline
\noindent 

Table \ref{tab:2_res} presents the further comparison of WavLM-Large and XLSR with different learning rate, and the final submitted fine-tuning system. 

\begin{table}[!htp]
\centering
\caption{Comparison of different learning rate and fine-tuning.}
\label{tab:2_res}
\begin{tabular}{ccccc}
\toprule
\textbf{SSL} & \textbf{Learning} &  \textbf{Fine} & \multicolumn{2}{c}{\textbf{EER} (\%)} \\
\cmidrule(l){4-5}
\textbf{Model} & \textbf{Rate} & \textbf{Tune} & \textbf{Adp2$^{\star}$} & \textbf{Test} \\
\midrule
WavLM-Large & 1e-6 & -          & 33.56 & 45.46 \\
WavLM-Large & 1e-7 & -          & 47.81 & -      \\
WavLM-Large & 1e-8 & -          & 48.57 & -      \\
XLSR       & 1e-6 & -          & 34.32 & 44.32 \\
XLSR       & 1e-7 & -          & 24.81 & -      \\
XLSR       & 1e-8 & -          & 18.16 & -      \\
\midrule
XLSR       & 1e-8 & Adp2$^{\star}$     & \textbf{1.33}  & \textbf{20.58} \\
\bottomrule
\end{tabular}
\vspace*{-4mm}
\end{table}

The XLSR model with $lr=1e-8$ yields relatively better performance among the six models without fine-tuning. Thus, we select XLSR with $lr=1e{-8}$ to do further fine-tune on  Adp2$^{\star}$ set. Consistent with the conclusion of LF track, the mismatches between training data and test data can be corrected by using fine-tuning strategy in PF track. The final submitted single XLSR model can achieve 1.33 \% EER on Adp2$^{\star}$ and 20.58\% EER on the evaluation set, respectively.

\section{Conclusion}

In this study, we focused on the LF and PF tracks in ADD 2022. For the LF track, we presented low-quality data augmentation, domain adaptation, and greedy fusion of various complementary features to capture spectro-temporal artifacts under noisy environments. We performed a series of ablation experiments to demonstrate the effectiveness of data augmentation and fine-tuning strategies. Moreover, we analyzed the clustering characteristics of subsystems, and explained the effectiveness of our proposed greedy fusion strategy through the visualization method. The submitted primary system reached the 25.91\% EER on the test1 set and ranked 4th. For the PF track, we investigated several popular SSL models to extract temporal information for partially fake utterances, and compared the effect of learning rate on the SSL model. The final single XLSR model achieves 20.58 \% EER, which ranked 5th on the test2 set. Our main contribution is an overall assessment of mainstream front-end features and back-end models on this new mandarin deep fake database. In the future, we will continue concentrating on eliminating scene mismatches and defending the spoofed attacks.


\begin{acks}
This work was supported by the National Natural Science Foundation of China under Grant 62176182, JST CREST Grants (JPMJCR18A6, JPMJCR20D3 and JPMJFS2136), MEXT KAKENHI Grant (21H04906), and the Agency for Science, Technology and Research (A*STAR), Singapore, through its Core Project Scheme (Project No. CR-2021-005).
\end{acks}

\bibliographystyle{IEEEbib_lk}
\bibliography{strings,refs}

\end{document}